\documentclass[aps,pra,twocolumn,amsmath,amssymb,superscriptaddress]{revtex4}

\usepackage{graphicx}
\usepackage{bm}
\usepackage{physics}
\usepackage{hyperref}

\begin{document}

\title{Mapping Between Nonlinear Sch\"odinger  Equations with Real and Complex Potentials}

\author{Mario Salerno} 
\affiliation{Dipartimento di Fisica
  ``E.R. Caianiello'', and INFN Sezione di Napoli - Gruppo Collegato di Salerno,
  Universit\`a di Salerno, Via Giovanni Paolo II, 84084 Fisciano (SA),
  Italy}

\begin{abstract}
A mapping between stationary solutions of nonlinear Sch\"odinger  equations
with real  and complex potentials is constructed and a set of exact  solutions with real energies are obtained for a large class of complex potentials.
As specific examples we consider the case of dissipative periodic soliton solutions of the nonlinear Schr\"odinger equation with complex potential.
\end{abstract}

\maketitle

\section{Introduction}
Nonlinear wave phenomena with time evolutions governed by non hermitian Hamiltonians  are presently attracting a great interest both from the theoretical and from the applicative point of view. The non hermiticity is in general due  to the presence of a complex potential in the Hamiltonian  accounting  for typical dissipative and amplification effects met in classical and quantum contexts~\cite{optics, QM}. In particular, dissipative solitons~\cite{LNP} of the nonlinear Schr\"odinger (NLS) equation with periodic complex potentials have been extensively investigated during the past years in connections with the propagation of light in nonlinear optical fibers with periodic modulations of the complex refractive index~\cite{Musslim1,Staliunas}. Recently similar studies were done for matter wave solitons of  Bose-Einstein condensates (BEC) trapped  in absorbing optical lattices~\cite{Abd08,Bludov10} and in the presence of three body interatomic interactions~\cite{AbdSal05}. In the linear context, the recent discovery~\cite{Bender} that  the Schr\"odinger eigenvalue problem  with complex potentials that are invariant under the combined  parity and time reversal symmetry (so called ${\cal PT}$-potentials), may have fully real spectrum, has raised interest also in view of possible connection with the theory of dissipative quantum systems~\cite{JPA}. Complex potentials with ${\cal PT}$-symmetry are presently investigated in nonlinear optics~\cite{Longhi} where it has been demonstrated  that nonlinear media with {\it linear} damping and amplifications that are  ${\cal PT}$-symmetric can support stable stationary localized and periodic states~\cite{Musslim2}. Also, quite recently,  physical systems with ${\cal PT}$-symmetry have been successfully implemented in real experiments~\cite{exp1,exp2,exp3}.
Solutions of the  NLS equation with a complex potential which belong to the real part of the spectrum  (real energies or real chemical potentials) can exist, however,  for generic complex potentials and it is therefore of interest to characterize them in general, independently from the ${\cal PT}$-symmetry.

The aim of the  present  paper is to show how one can  systematically construct stationary solutions of the  complex nonlinear Schr\"odinger equation via  a mapping between real and complex NLS equations. The problem is formulated  in terms of a nonlocal  eigenvalue problem which involves  only real potentials, whose  eigenfunctions and eigenvalues  fix amplitudes and energies  of  the stationary solutions of the complex  NLS equation,  respectively. The complex potentials and the phases of the solutions are also determined self-consistently through the mapping. We illustrate our approach on the example  of the NLS equation with different complex potentials for which we construct periodic dissipative solitons in the form of elliptic functions.

The paper is organized as follows. In section II we introduce model equations and illustrate the mapping  used to determine the solutions. In section III we show how to construct exact  solutions of the  NLS with periodic complex potentials. In the last section  the main results of the paper will be briefly summarized.

\section{Model Equations and Mapping }
The model equation we consider is the NLS equation with real and complex potentials
both of linear and nonlinear types, e.g.
\begin{equation}\label{sys}
i \psi_t=-\frac{1}{2}\psi_{xx}+
(V_l(x)+iW_l(x))\psi + (\sigma + V_{nl}(x)+iW_{nl}(x))|\psi|^2 \psi.
\end{equation}
The case of the linear Schrodinger equation (e.g. $\sigma=V_{nl}=W_{nl}=0$) can be used as an example of quantum dissipative system. In the nonlinear case the above  equation can appear in connection with several interesting phenomena including  light propagation in photonic crystals and Bose-Einstein condensates. Due to the possibility of  different physical applications we shall keep Eq. (\ref{sys}) in normalized form, looking for stationary solutions of the type
\begin{equation}
\psi(x,t) =  A(x) e^{i \theta(x)} e^{-i \omega t}
\label{stationary}
\end{equation}
with the amplitude $A(x)$ and phase $\theta(x)$ as real functions. Substituting this expression into Eq.(\ref{sys}), we obtain the system of equations
\begin{eqnarray}
\label{sys2}
\omega A + \frac{1}{2}A_{xx}- \sigma A^3 - \frac{A}{2}(\theta_x)^2 -V_l A - V_{nl}A^3 = 0\\
\frac 12  A\theta_{xx} + A_x\theta_x  - W_l A - W_{nl} A^3 = 0.
\label{sys3}
\end{eqnarray}
These equations can be easily separated. In this respect notice that by multiplying Eq. (\ref{sys3}) by $A$ and integrating it twice one obtains
\begin{equation}
\frac 12 \theta(x)= B_2 + \int_{-\infty}^x \frac{ B_1 + F(y)}{A^2(y)}dy
\label{tetagen}
\end{equation}
with
\begin{equation}
\label{effe}
F(y)= \int_{-\infty}^y  \left[ W_l(z)  + W_{nl}(z) A^2(z) \right] A^2(z) dz,
\end{equation}
and $B_1, B_2$ integration constants. By substituting Eq. (\ref{tetagen}) into Eq. (\ref{sys2}) we obtain  the following nonlinear  eigenvalue problem for the real amplitude $A$:
\begin{equation}
\label{eigenval-prob-gen}
\left\{-\frac 12 \frac{\partial}{\partial x^2} + V_l + (\sigma + V_{nl}) A^2 + 2 \left(\frac{F(x)}{A^2} \right)^2 \right\} A = \omega A
\end{equation}
where the integration constants $B_1, B_2,$ have been  fixed to zero for simplicity.
Note that for stationary solutions Eq. (\ref{eigenval-prob-gen}) is completely  equivalent to  Eq. (\ref{sys}) in the sense that any solution of (\ref{eigenval-prob-gen}) gives a stationary solution of (\ref{sys}) with the phase fixed  by (\ref{tetagen}). Also note that  the dependence on the complex potentials in the eigenvalue problems comes through the function $F$ and for an arbitrary $F(x)$ (e.g. arbitrary  complex potentials) the problem can become singular.
It is possible, however,  to construct potentials $W_l$ and $W_{nl}$ (e.g. functions $F$)
so that the solutions of (\ref{eigenval-prob-gen}) are regular. This establishes a mapping
between stationary solutions of the NLS equation with real potentials and stationary solution of Eq. (\ref{sys}) with the phase given by (\ref{tetagen}). In this respect, one can take  $F$  in general to be an analytical function of $A^2$ and derivatives e.g. $F(x) \equiv F(A^2, (A^2)_x,...)$. In the simplest case $F$ can be taken of the form
\begin{equation}
F(x)=\frac{1}{2} {C_n} A^{n+2},\;\;\;\; n=0,1,2...
\label{complex}
\end{equation}
with $C_n$ constants to be determined.
Equation (\ref{eigenval-prob-gen}) then reduces to the following NLS real  eigenvalue problem:
\begin{equation}
\left\{-\frac 12 \frac{\partial}{\partial x^2} + V_l + (\sigma  + V_{nl}) A^2 + \frac{C_n^2}{2} A^{2n} \right \} A = \omega A
\label{eigen-prob-real}
\end{equation}
which can be solved analytically for particular forms of the potentials $V_l$,  $V_{nl}$, or numerically with high accuracy (using for example the self-consistent method discussed in~\cite{ms05}) for generic real potentials. In the following we therefore assume that
the real amplitudes $A$ and frequencies $\omega$ for given  $V_l$ and $V_{nl}$ are exactly obtained  from (\ref{eigen-prob-real}), either analytically or numerically.

On the other hand from  Eq. (\ref{complex}) one can characterize the complex potentials which support such solutions.
Using Eq. (\ref{effe}) we have indeed that Eq. (\ref{complex}) is satisfied if the amplitude $A$ is related to $W_l$ and $W_{nl}$ by the relation
\begin{equation}
W_l + W_{nl}A^2= C_n (\frac {n+2}{2 n}) \frac{d A^n}{d x}
\label{wl}
\end{equation}
and from Eqs. (\ref{tetagen}), (\ref{complex}), one gets that the phase is given by
\begin{equation}
\theta(x) =  C_n \int_{-\infty}^{x} {A^{n}}.
\label{teta}
\end{equation}

Note that in this case Eq. (\ref{wl}) allows to relate the constant $C_n$ to the amplitude of the solution, $A_0$, and the amplitudes $W_{0l}$, $W_{0nl}$, of the
linear and nonlinear complex potentials, respectively. In particular, for the case $W_{nl}=0$ we have that
\begin{equation}
C_n= \frac {2}{n+2}  \frac{W_{0l}}{A_0^{n}},\;\;\;\; W_{0nl}=0
\label{cost-wnl0}
\end{equation}
while for $W_{l}=0$ one obtains
\begin{equation}
C_n= \frac {2}{n+2}  \frac{W_{0nl}}{A_0^{n-2}},\;\;\;\; W_{0l}=0.
\label{cost-wl0}
\end{equation}
It is worth to note that while the case $n=1$ leads  to a pure cubic NLS eigenvalue problem, the case $n>1$  introduces higher order
nonlinearities in Eq. (\ref{eigen-prob-real}) which  can however be eliminated by redefying the linear real potential as
\begin{equation}
V_{l} = \tilde V_l - \frac{{C_n}^2}{2}  A^{2n},
\label{lin-n2}
\end{equation}
or the nonlinear real potential as
\begin{equation}
V_{nl} = \tilde V_{nl}- \frac{{C_n}^2}{2} A^{2n-2}
\label{nlin-n2}
\end{equation}
(or a combination of both).
Also notice that
Eqs. (\ref{stationary}), (\ref{wl}) - (\ref{cost-wl0}) allow to map solutions of the real eigenvalue problem (\ref{eigen-prob-real}) into solutions of the NLS equation  (\ref{sys}) with
the corresponding complex potentials determined as in (\ref{wl}).
It is clear that this approach can be extended to functions of the of the type
\begin{equation}
F(x)=\frac{1}{2} \sum_{n=0}^k{C_n} A^{n+2},\;\;\;\; k=0,1,2...
\label{complexg}
\end{equation}
In this case coefficient $C_n$ are self-consistently determined from the real eigenvalue problem
\begin{equation}
\left\{-\frac 12 \frac{\partial}{\partial x^2} + V_l + (\sigma  + V_{nl}) A^2 + \frac 12 (\sum_n {C_n} A^{n})^2 \right \} A = \omega A
\label{eigen-prob-realg}
\end{equation}
and complex potentials and phase are given by
\begin{equation}
W_l + W_{nl}A^2=\frac{1}{A^2} \frac{d F}{dx}
\label{wlg}
\end{equation}
\begin{equation}
\theta(x) =\sum_n C_n \int_{-\infty}^{x} {A^{n}}.
\label{tetag}
\end{equation}
Note that the sum in Eq. (\ref{complexg}) can include infinite terms and to have a map between real and complex  NLS equations it is necessary
to subtract higher order nonlinearities from the real linear and nonlinear potentials as done in  Eqs. (\ref{lin-n2})-(\ref{nlin-n2}).
Finally we remark that if the functions  $A_x/A, A_{xx}/A, ... $ are bounded, the expression (\ref{complexg}) can be further generalized as
\begin{equation}
F(x)=\sum_{n,m}^k{C_{n,m}} \frac {d^m A^{n+2}}{dx^m}
\label{complexgg}
\end{equation}
with $C_{n,m}$ suitable constants and with the complex potentials determined as (\ref{wlg}) .
In all these cases
a map between solutions of the real eigenvalue problem (\ref{eigen-prob-realg}) and  solutions of the NLS equation  (\ref{sys}) is constructed.
The mapping guarantees that the constructed solutions always have real energies and can be therefore of physical interest. In the following we illustrate how the mapping works on some specific example.

We finally remark that a similar approach based on a priori fixing of  the solution and a posteriori determination of the complex potential, has been considered also in~\cite{BK, AKSY}, although not in terms of a mapping between stationary solutions of  NLS equations. In the following we illustratehow the mapping works on some specific example.
\section{Nonlinear Schr\"odinger Equation with Complex Potentials}

\subsection{\bf Case n=1}
Let us consider first  the simplest ansatz (\ref{complex}) with  $n=1$. We fix the
nonlinearity to be attractive ($\sigma<0$) and restrict to linear complex potentials  (i.e. $W_{nl}= V_{nl}=0$) and with linear potential of the form $V_l= V_{0l} \, \mbox{cn}^2(x,k)$. In this case the real eigenvalue
problem (\ref{eigen-prob-real})
\begin{equation}
\left[-\frac 12 \frac{\partial}{\partial x^2} + V_{0l}\; \mbox{cn}(x,k)^2 + (\sigma  +
\frac{C_1^2}{2}) A^2 \right] A = \omega A.
\end{equation}
admits the following exact solutions in terms of elliptic functions
\begin{eqnarray}
a)\;\;\;\; A(x)&=&A_0 \, \mbox{cn}(x,k),\;\; A_0=\pm\sqrt{\frac{2(k^2+V_{0l})}{2 |\sigma| - C_1^2}},\;\; \nonumber\\
&& \omega =\frac{1- 2 k^2}{2} \\
b)\;\;\;\;A(x)&=&A_0 \, \mbox{sn}(x,k), \;\;A_0=\pm \sqrt{\frac{2(k^2+V_{0l})}{C_1^2- 2 |\sigma|}},\nonumber\\
&& \omega =\frac{1+k^2}{2} + V_{0l} , \\
c)\;\;\;\;A(x)&=&A_0\, \mbox{dn}(x,k), \;\; A_0=\pm\frac{1}{k}\sqrt{\frac{2(k^2+V_{0l})}{2|\sigma|-C_1^2}},\nonumber\\
&& \omega =\frac{k^2}{2}-1+V_{0l}(1-\frac{1}{k^2}).
\label{solution-cn2}
\end{eqnarray}
Similar solutions can be constructed for the case of a repulsive nonlinearity $\sigma>0$ with
linear potentials of the form $V_l= V_{0l} \, \mbox{sn}^2(x,k)$. In this case we have
\begin{eqnarray}
d)\;\;\;\; A(x)&=&A_0 \, \mbox{cn}(x,k), \;\; A_0=\pm\sqrt{\frac{2(V_{0l}-k^2)}{C_1^2 + 2 \sigma }},\;\; \nonumber\\
&& \omega = \frac{1- 2 k^2}{2} +V_{0l}\\
e)\;\;\;\;A(x)&=&A_0 \, \mbox{sn}(x,k),  \;\;A_0=\pm \sqrt{\frac{2(k^2-V_{0l})}{C_1^2 + 2 \sigma}},\nonumber\\
&& \omega = \frac{1+k^2}{2}, \\
f)\;\;\;\;A(x)&=&A_0\, \mbox{dn}(x,k),  \;\; A_0=\pm\frac{1}{k}\sqrt{\frac{V_{0l}-k^2}{C_1^2 + 2\sigma}},\nonumber\\
&& \omega = \frac{k^2}{2}-1+\frac{V_{0l}}{k^2}.
\label{solution-sn2}
\end{eqnarray}

Using the above mapping we can readily construct the stationary solutions of the corresponding complex NLS with
linear complex potentials given by
\begin{equation}
 W_l= \frac 32 C_1 A_x, \quad\quad C_1 = \frac 23 \frac{W_{0l}} {A_0}
\end{equation}
and with the phase given by $\theta(x)= C_1 \int_{-\infty}^x A(y)dy$.
Thus, for example,  from the solution a) we get
\begin{eqnarray}
&&A = A_0 \mbox{cn}(x,k),\;\;V_l(x)=V_{0l}\, \mbox{cn}^2(x,k), \\
&& A_0=\frac{\sqrt{9 \left(k^2 + {V_{0l}}\right)+2
   {W_{0l}}^2}}{3 \sqrt{|\sigma|}},\;\;\;\omega = \frac{1-2k^2}{2}, \nonumber\\
&&W_l=-W_{0l}\, \mbox{sn}(x)\mbox{dn}(x), \nonumber \\
&&\theta(x)=\frac{2W_{0l}}{3k}\mbox{arccos}(\mbox{dn}(x)). \nonumber
\end{eqnarray}
In  similar manner one proceeds with the other solutions above.
It is also clear that exact solutions of this type can be constructed also for other types of linear
elliptic potentials (we omit them for brevity).

\subsection{\bf Case n=2}
As a further application  of the ansatz (\ref{complex}) we consider the case $n=2$ for which the mappings
involves higher order nonlinearities. We assume as before that
$V_{nl} = W_{nl} = 0$.  In order to balance the quintic nonlinearity in
Eq. (\ref{eigen-prob-real}),
the potential $V_l$ must be taken as in Eq. (\ref{lin-n2}). We take $\tilde V_l= V_{0l} \, cn^2(x,k)$ and consider
a solution of the form $A(x) = A_0 \, \mbox{cn}(x,k)$. One can then check that this is a
solution of (\ref{eigen-prob-real}) with
\begin{equation}
V_l(x)= V_{0l}\mbox{cn}^2(x,k) - \frac{C_2^2}{2} A_0^4 \,\mbox{cn}^4(x,k).
\end{equation}
if
$A_0^2 = V_{0l} + k^2, \;\;\; \omega = \frac{1-2k^2}{2}$
From the mapping we then have that $C_2=\frac{W_{0l}}{2 A_0^2}=\frac{W_{0l}}{2(V_0 + k^2)}$,
\begin{equation}
W_l(x)= 2 C_2 A A_x = - W_{0l}\, \mbox{cn}(x) \mbox{sn}(x) \mbox{dn}(x),
\end{equation}
and the phase is
\begin{equation}
\theta(x)= x - \frac{x}{k^2} + \frac{1-k^2 + k^2 \mbox{cn}^2(x,k)}{k^2 \mbox{dn}^2(x,k)}E(\mbox{am}(x,k),k).
\end{equation}

As a further example of $n=2$ we consider  the case of pure nonlinear optical lattices, i.e. $V_l = W_l = 0$.
Fixing $\tilde V_{nl}=0$ and looking for solutions of the type $A(x)= A_0 \mbox{cn}(x,k)$, we have from
Eq. (\ref{nlin-n2}) that
that
\begin{equation}
V_{nl}(x)= - \frac{C_2^2}{2} A_0^2 \,\mbox{cn}^2(x,k).
\end{equation}
with $C_2$ fixed according to Eq. (\ref{cost-wl0}) as $C_2 = W_{0nl}/2$.
One can easily check that this is indeed a solution of the eigenvalue problem (\ref{eigen-prob-real})
if
\begin{equation}
\omega=\frac 12 - k^2, \;\;\; A_0=\frac{k}{\sqrt{|\sigma|}}
\end{equation}
(we consider $\sigma<0$). From the mapping  we have that this is also a solution of
the NLS with the complex part of the nonlinear potential
fixed according to Eq. (\ref{wl}) as
\begin{equation}
W_{nl}(x)= 2 C_2 \frac{A_x}{A}= - W_{0nl} \frac{\mbox{sn}(x,k)\mbox{dn}(x,k)}{\mbox{cn}(x,k)}.
\end{equation}

For the cases $n>2$ one can proceed in similar manner.

\subsection{\bf General case}
Let us now consider an example with the more general ansatz (\ref{complexg}). To this regard we take
$F(x)= \frac 12 (C_0 + C_2 A^2) A^2$ and look for solutions of the form
$A(x)= A_0 \mbox{dn}(x,k)$. Let us fix the linear potentials as $V_l = W_l=0$ and the real nonlinear potential as
$V_{nl}= V_{0nl} -{\alpha_2^2} A^2$ with $V_{0nl}$ a constant and with $\alpha_n=\frac{C_n}{\sqrt{2}}$, $n=0,2$
(notice that we fixed all coefficients for $n\ne 0,2,$ equal to zero).
By substituting these expressions into the real eigenvalue problem we find that $A(x)$ is indeed a solution if
\begin{equation}
\omega=\alpha_0^2 + \frac{k^2}{2}-1,\quad\quad \alpha_0 = -\frac{1 + A_0^2 (\sigma + V_{0nl})}{2 A_0^2 \alpha_2}.
\label{om}
\end{equation}
Thus, for example, if we fix $V_{0nl}=2/k^2$, $\alpha_2=-1/k$ and consider $\sigma=1$ (repulsive interactions),  we have
\begin{equation}
V_{nl}=\frac{2- \mbox{dn}^2(x,k)}{k^2}=\frac{1}{k^2}+\mbox{sn}^2(x,k)).
\end{equation}
From Eq. (\ref{om}) we have
\begin{eqnarray}
&& \omega=\sigma -1 +\frac{1}{A_0^2}+\frac{k^2  \left[(1+ 2 \sigma  A_0^2)+
   (\sigma^2+2)A_0^4\right] }{4
   A_0^4}+\frac{1}{k^2} \nonumber \\
&&
\alpha_0 = \frac 1k + \frac{ k (1 + \sigma A_0^2)}{2 A_0^2},
\end{eqnarray}
and from (\ref{complexg}) we get the  function $F$ as
\begin{eqnarray}
F(x) &=& \frac{1}{\sqrt{2}}(\alpha_0 + \alpha_2 A^2) A^2 = \nonumber \\ &&
\frac{A_0^2 \mbox{dn}^2(x,k)}{2 \sqrt{2} k}\left[ 2+\frac{k^2}{A_0^2}(1+ \sigma A_0^2)-2 A_0^2 \mbox{dn}^2(x,k)\right] \nonumber
\end{eqnarray}
Substituting into Eqs. (\ref{wlg}) we finally  get the complex potential as
\begin{eqnarray}
W_{nl} &=& \sqrt{2}k \frac{\mbox{sn}(x,k)\mbox{cn}(x,k)}{\mbox{dn}^3(x,k)} \times  \nonumber \\
&&
\left[2 -\frac{1}{A_0^2} - \frac{k^2}{2A_0^4}
(1+ \sigma A_0^2)-2 k^2 \mbox{sn}^2 (x,k)  \right].
\end{eqnarray}
The phase of the solution can be readily obtained from Eq. (\ref{tetag}). Notice that in the case $\sigma=1, A_0=1$,
this solution coincides with the one derived  in~\cite{AKSY} with a slightly different approach.
  We remark that the above solutions of the complex NLS equations not only have  real energies  (chemical potentials) but are also  found to be stable (not shown here for brevity) under time evolution.

\section{Conclusions} In conclusion we have shown the possibility to  construct stationary solutions of the nonlinear Schrodinger equation with complex potentials via a mapping with stationary solutions  of the NLS equation with suitable real potentials. In particular, we showed that by means of the  mapping it is possible to construct sets of exact  solutions with real energies for different types of complex potentials. The presented  approach can be applied also to other types of equations, including  the linear  Schr\"odinger equation describing quantum  dissipative oscillators, and the NLS equation with  arbitrary higher order nonlinearities, as it will be discussed elsewhere.

\section*{Acknowledgements}
This paper contributes to the special issue of the {\it Journal of Geometry and Symmetry in Physics}, JGSP {\bf 32} 25-35 (2013), in honor of the 65th birthday of Prof. Vladimir Gerdjikov. Partial support from the Ministero dell' Istruzione, dell' Universit\`{a} e della Ricerca (MIUR) through a \textit{Programma di Ricerca Scientifica di Rilevante Interesse Nazionale} (PRIN) 2010-2011 initiative, is acknowledged.



\begin{thebibliography}{99}\itemsep=-.2pc


\bibitem{optics} I. Aranson and L. Kramer, {\it The World of the Complex Ginzburg-Landau Equation,} Rev. Mod. Phys. {\bf 74} (2002) 99-143.

\bibitem{QM} J. Muga, J. Palao, B. Navarro, I . Egusquiza, {\it Complex Absorbing Potentials,} Phys. Rep. {\bf 395} (2004) 357-426.


\bibitem{LNP}N. Akhmediev and A. Ankiewicz (Eds.) {\it Dissipative solitons,} Lecture Notes in Physics {\bf 661}, Springer-Velag, Berlin, (2005).

\bibitem{Musslim1} Z. Musslimani, K. Markis, R.El-Ganainy, and D. Christodoulides, {\it Optical Solitons in PT Periodic Potentials,}  Phys.Rev.Lett. {\bf 100} (2008) 030402-1-4.


\bibitem{Staliunas} K. Staliunas, R. Herrero, R. Vilaseca, {\it Subdiffraction and Spatial Filtering due to Periodic Spatial Modulation of the Gain-Loss Profile,} Phys. Rev.A {\bf 80} (2009) 013821-1-6.

\bibitem{Abd08}
F. Abdullaev, A. Gammal, H. da Luz., and L. Tomio, {\it Dissipative Dynamics of Matter-Wave Solitons in a Nonlinear Optical Lattice,}  Phys.Rev. A {\bf 76} (2007) 043611-1-9.

\bibitem{Bludov10}
Yu. Bludov and V. Konotop, {\it Nonlinear Patterns in Bose-Einstein Condensates in Dissipative Optical Lattices,}  Phys.Rev. A {\bf 81} (2010) 013625-1-8.

\bibitem{AbdSal05}F. Abdullaev and M. Salerno, {\it Gap-Townes Solitons and Localized Excitations in Low-Dimensional Bose-Einstein Condensates in Optical Lattices,}  Phys.Rev. A {\bf 72} (2005) 033617-1-12.

\bibitem{Bender}
C. Bender, S. Boettcher, {\it Real Spectra in Non-Hermitian Hamiltonians Having PT Symmetry,}  Phys.Rev.Lett. {\bf 80} (1998) 5243-5246.

\bibitem{JPA} see special issue no. 32: {\it The Physics of Non-Hermitian Operators}, J. Phys. A, {\bf 39} (2006).

\bibitem{Longhi}
S. Longhi, {\it Bloch Oscillations in Complex Crystals with PT Symmetry,}  Phys.Rev.Lett. {\bf 103} (2009) 123601-1-4.

\bibitem{Musslim2}
Z. Musslimani, K. Markis, R. El-Ganainy, and D. Christodoulides,  {\it Analytical Solutions to a Class of Nonlinear Schrodinger Equations with PT-Like Potentials,}   J.Phys. A {\bf 41} (2008) 244019.

\bibitem{exp1}A. Guo, G. Salamo, D. Duchesne, R. Morandotti, M. Volatier-Ravat, V. Aimez,
G. Siviloglou, and D. Christodoulides, {\it Observation of PT-Symmetry Breaking in Complex Optical Potentials,}  Phys. Rev. Lett. {\bf 103} (2009)  093902-1-4.

\bibitem{exp2} C. R¨uter,K. Makris, R. El-Ganainy, D. Christodoulides,
M. Segev, and D. Kip, {\it Observation of Parity-Time Symmetry in Optics,}  Nature Phys. {\bf 6} (2010) 192-195.

\bibitem{exp3} J. Schindler, A. Li, M. Zheng, F. Ellis, and T. Kottos, {\it Experimental study of active LRC circuits with PT symmetries,}
Phys. Rev. A {\bf 84} (2011) 040101-1-5.

\bibitem{ms05}
Mario Salerno, Laser Physics, {\it Macroscopic Bound States and the Josephson Effect
in Bose Einstein Condensates in Optical Lattices,}  {\bf 14} (2005) pp. 620-625.

\bibitem{BK}
V. Brazhnyi  and V. Konotop, {\it Theory of Nonlinear Matter Waves in Optical Lattices,}   Mod. Phys. Lett. B {\bf 18} (2004) 627-651.

\bibitem{AKSY} F. Abdullaev, V. Konotop,  M. Salerno, A. Yulin, {\it Dissipative Periodic Waves, Solitons, and Breathers of the Nonlinear Schrödinger Equation
with Complex Potentials,}  Phys. Rev. E {\bf 82}  (2010) 056606-1-6.

\end{thebibliography}
\end{document}